\documentclass[preprint,11pt]{elsarticle}

\usepackage{graphicx}
\usepackage{amssymb}

\usepackage{lineno}

\usepackage{mathtools}
\usepackage{amssymb}
\usepackage{algorithmic}
\usepackage{array}
\usepackage{mdwmath}
\usepackage{mdwtab}
\usepackage{eqparbox}
\usepackage{fixltx2e}
\usepackage{stfloats}
\usepackage{epsfig}
\usepackage{graphicx}
\usepackage[outdir=./]{epstopdf}
\usepackage[inline]{enumitem}
\usepackage{caption}
\usepackage{morefloats}
\usepackage{float}
\usepackage{multirow}
\usepackage{color}
\usepackage{subcaption}
\usepackage{amssymb}
\usepackage{color}
\usepackage{siunitx}
\usepackage{tikz}
\usetikzlibrary{shapes,arrows,decorations.markings}

\journal{Physical Communication}

\begin{document}

\begin{frontmatter}


\title{A 3D Human Body Blockage Model for Outdoor Millimeter-Wave Cellular Communication}



\author[tukl]{Bin Han}
 \ead{binhan@eit.uni-kl.de}
\author[huawei]{Longbao Wang}
 \ead{wanglongbao@huawei.com}
\author[tukl]{Hans D. Schotten}
 \ead{schotten@eit.uni-kl.de}

\address[tukl]{University of Kaiserslautern\\
	Institute for Wireless Communication and Navigation\\
	67663 Kaiserslautern, Germany}
\address[huawei]{HUAWEI R\&D Center\\
	No. 2222, Xin Jin Qiao Road,	Pudong, Shanghai, China}

\begin{abstract}
Blocking is one of the most important challenges in exploiting millimeter-wave for fifth-generation (5G) cellular communication systems. Compared to blockages caused by buildings or terrains, human body blockage exhibits a higher complexity due to the mobility and dynamic statistics of humans. To support development of outdoor millimeter-wave cellular systems, in this paper we present a novel 3D physical model of human body blockage. Based on the proposed model, the impact of human body blockage on frame-based data transmission is discussed, with respect to the system specifications and environment conditions.
\end{abstract}


\end{frontmatter}


\section{Introduction}\label{sec:intro}
The demand of human society for data traffic has been explosively growing since decades. Beginning with 2.4 kpbs in the first generation (1G) systems in 1980s, the data rate of wireless technologies has increased to several Gbps in the current Long Term Evolution Advanced (LTE-A) systems. Correspondingly, the exploited bandwidth has also increased from \SI{30}{\kilo\hertz} to \SI{20}{\mega\hertz}. In future, the fifth generation (5G) technologies on horizon are expected to provide over 50 Gbps data rate with new bands up to \SI{60}{\giga\hertz} by the year of 2020 \cite{gupta2015survey,soldani2015horizon}. To quench such a thirst of bandwidth, the interest in using millimeter-wave (mmWave) frequencies for cellular communications has kept rising since the beginning of this decade \cite{pi2011introduction}. Today, mmWave communication is widely considered as an essential part of the upcoming 5G technologies.

Compared to legacy wireless technologies at lower radio frequencies, mmWave systems generally suffer from significantly higher path loss and penetration loss. To compensate such losses, highly directive antenna arrays are now widely considered as essential in mmWave systems, to provide enough channel budget for a reliable transmission. This results in a significant difference in the signal-to-noise ratio (SNR) between line-of-sight (LOS) and non-line-of-sight (NLOS) channels, especially in outdoor scenarios, where few reflectors are available near the transmitter and the receiver. Meanwhile, due to the short wavelength, mmWave transmission can be easily shadowed or blocked by objects of small size, such as human bodies or vehicles. Therefore, as pointed out by \textit{Niu} et al. \cite{niu2015survey} and \textit{Andrews} et al. \cite{andrews2017modeling}, blocking is one of the most important traits of mmWave cellular systems.

Blockages in cellular communications can be generally classified into two categories, according to the blocking object:
\begin{enumerate}
	\item macroscopic rigid obstructions such as buildings or terrains; 
	\item obstacles of intermediate scales (still significantly larger than the wavelength), such as trees, human bodies and vehicles.
\end{enumerate} 
Plenty mature models are available for the first class, e.g. the 3rd Generation Partnership Project (3GPP) model for incorporating blockages \cite{3gpp2017further},  the random shape theory model \cite{haenggi2014mean}, the LOS ball model \cite{bai2015coverage} and the Poisson line model \cite{baccelli2015correlated}. However, the efforts in modeling the latter type of blockages, especially human body blockages (HBBs), are still fresh but limited. It was reported by \textit{Collonge} et al. \cite{collonge2004influence} that the SNR of mmWave systems can be reduced by tens of \SI{}{\dB}s, when the LOS path is blocked by a person - either the device user itself or a pedestrian. This penetration loss through human bodies was afterwards confirmed and further studied by \textit{Lu} et al. \cite{lu2012modeling} and \textit{Ragagopal} et al. \cite{rajagopal2012channel}, known to float from \SIrange{20}{40}{\dB}.  Focusing on the self-blockage events, a 2D blocking model was proposed by \textit{Bai} and \textit{Heath, Jr.} \cite{bai2014analysis}, where the impact of elevation angle of transmitter to receiver is ignored. For pedestrian blockages, \textit{Venugopal} et al. proposed a blocking model in scope of indoor wearable device networks, approximating every human body as a cylinder \cite{venugopal2015analysis,venugopal2016millimeter}. A similar geometry model was applied by \textit{Gapeyenko} et al. \cite{gapeyenko2016analysis} on urban outdoor scenarios, extended with stochastic approaches to model the pedestrian blockage probability as function of transmitter-receiver locations / dimensions and human densities / dimensions. In \cite{3gpp2017study}, the 3GPP provided two different blocking models for millimeter wave propagation, covering human body and vehicle blockages.

Compared to building and terrain blockages, HBB shows a more complex characteristic. First, it is more dynamic due to the mobilities of both the user and the pedestrians. Second, it is sensitive to the pedestrian density, which can strongly vary between different places or time periods. Furthermore, a human is short enough to keep a LOS path available over its head, even when it is standing between the transmitter and the receiver in overlooking view. This phenomena gives HBB a high dependency on the system deployment, especially on the height of antenna installation. Yet most current available models are either two-dimensional \cite{bai2014analysis,venugopal2015analysis}, or specified to some particular system deployment \cite{venugopal2016millimeter}, or even empirically instead of physically built \cite{3gpp2017study}. Therefore, they cannot fulfill the requirements of flexibly modeling HBB events in variant outdoor scenarios. The statistic geometry model in \cite{gapeyenko2016analysis} has provided a satisfying characterization of pedestrian blockage probability in outdoor scenarios, given a substantial analysis on the physical channel. Nevertheless, it is still available for further improvement with concerns about the blockages' impact on the performance of data transmission.

In this work, we try to propose a novel 3D physical model of HBB to support modeling mmWave channels, and to study the impact of HBB on outdoor mmWave communications. The paper is organized as follows. In Sec.\ref{sec:sys_model} we build a simple and general system model for a clear problem description. Then we propose our 3D HBB model in Sec.\ref{sec:3d_hbb_model}, including a self-blockage model and a pedestrian blockage model. Afterwards, in Sec.\ref{sec:analysis} we set up a simple single-user sidewalk scenario, and investigate how HBB may impact the transmission efficiency with respect to the pedestrian density, the user position and the system specifications. Both analytical discussions and numerical computations are presented. At the end we close this paper with our conclusions and some outlooks in Sec.\ref{sec:conclusion}.

\section{System Model}\label{sec:sys_model}
\subsection{HBB in Outdoor mmWave Communication Scenarios}
\begin{figure}[!h]
	\centering
	\includegraphics[width=.65\textwidth]{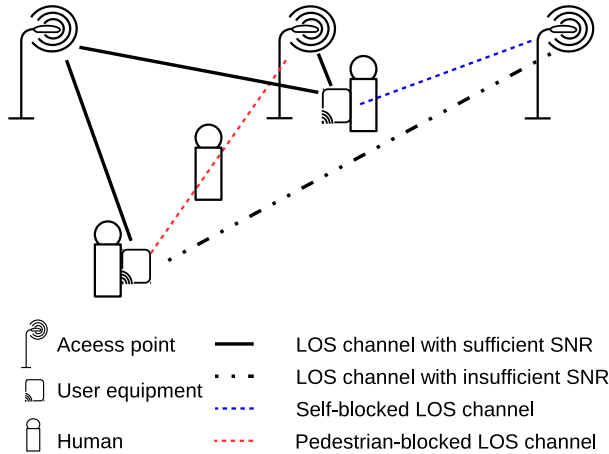}
	\caption{High-speed data transmission over millimeter wave can only take place over unblocked LOS channels with satisfying SNR.}
	\label{fig:data_shower}
\end{figure}

The HBB phenomenon in outdoor mmWave communication scenarios can be briefly illustrated by Fig. \ref{fig:data_shower}. Multiple mmWave access points (APs) can be available in a relatively small service area to guarantee the coverage and service availability. There are people with or without communicational user equipments (UEs) moving at low velocities in the service area. For convenience, in this paper we use the term \textit{users} to distinguish the humans who are carrying UEs from those without UE, which are referred to as \textit{pedestrians}. The LOS channel between a UE and an AP can be blocked by the user itself or a pedestrian. Even when a LOS channel is not blocked, it may fail to provide a sufficient signal-to-noise ratio (SNR) due to a strong path loss over long distance between UE and AP. Only LOS channels over short distances can be deployed for a satisfying high-speed transmission.

\subsection{Blockage-Caused Frame Loss}

\begin{figure}[!h]
	\centering
	\includegraphics[width=.7\textwidth]{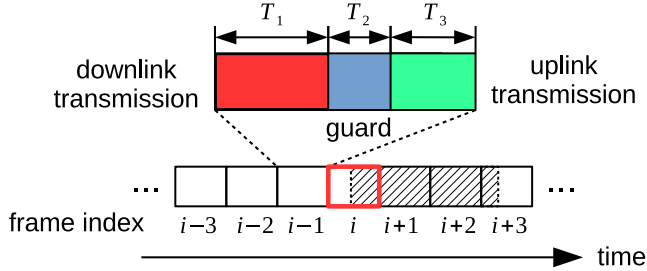}
	\caption{HBB causes failures in frame-based transmission. The shadow illustrates the duration of a human body blockage; the start of the frame loss is highlighted in red lines.}
	\label{fig:frame_failure}
\end{figure}

Most modern cellular communication systems, if not all, use synchronized frame-based transmission protocols. Under such protocols, data are synchronously transmitted in separated time frames of certain length. In this work we consider a Time-Division-Duplex (TDD) system, in which the frames are structured as illustrated in Fig. \ref{fig:frame_failure}. Each frame consists of three slots, for downlink (DL) transmission, guarding and uplink (UL) transmission, respectively. Both DL and UL pilots are transmitted in the guard slot to estimate the channel state information (CSI). Let $T_1$, $T_2$ and $T_3$ denote the lengths of a DL transmission slot, of a guard slot and of a UL transmission slot, respectively, we have the frame length $T=T_1+T_2+T_3$. 

Depending on the channel coherence time and the use case, $T_1$, $T_2$ and $T_3$ can be flexibly configured, but generally there is a minimal value for the length of each slot, known as the transmission time interval (TTI). In legacy LTE-A systems, this value is \SI{1}{\milli\second}. Considering the fact that the Doppler shift increases proportionally to the carrier frequency, mmWave systems generally suffer from shorter channel coherence time in comparison to LTE-A systems, and therefore require shorter delay in transmission and feedback. Concerning of this, it has been recently discussed and considered by the 3GPP in scope of \textit{New Radio} (NR) technologies, to employ shortened TTI down to a \textit{short TTI} (sTTI) of 2 symbols i.e. \SI{143}{\micro\second} in the future \cite{huawei2016overview,erricson2016system}. Meanwhile, according to the measurements reported in \cite{collonge2004influence}, the duration of a single human body blockage event typically ranges from \SIrange{0.5}{2}{\second}, which is significantly larger than the standard LTE-A TTI and sTTI. In other words, a blockage event generally lasts over multiple time frames. Once a HBB occurs during some time frame $i$, it can be recognized by the AP after the UL transmission slot, so that the AP can reallocate its radio resources for other unblocked UEs in the next frames ($i+1$ to $i+3$). Meanwhile, the transmission in frame $i$ is failed, generating no effective throughput. Note that the HBB may occur in the guard or UL transmission slot, leaving the DL transmission undisturbed. Nevertheless, due to the failed UL transmission the AP cannot be reported about the successful DL transmission, and has to retransmit this DL slot again with the next opportunity. Therefore, each HBB event leads to a data/resource loss of one time frame.

\subsection{Coordinate System}
\begin{figure}[!h]
	\centering
	\includegraphics[width=.5\textwidth]{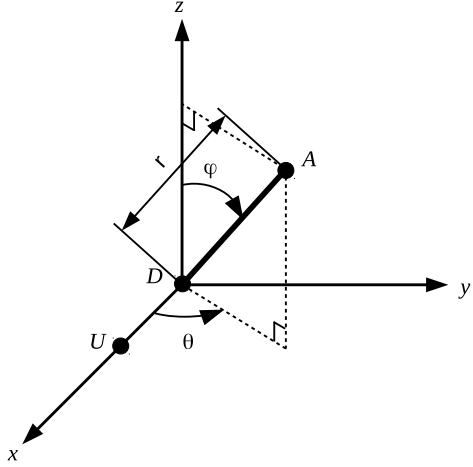}
	\caption{The UE-AP relative position can be presented in spherical coordinates as $[\theta,\varphi,r]$. Note that the $x$ axis here is in the direction from the UE to the user body $U$, and the $z$ axis points vertically upward.}
	\label{fig:coordinate_system}
\end{figure}

For a convenient spatial description in the rest part of the paper, here we use the spherical coordinate as illustrated in Fig.\ref{fig:coordinate_system}. The position of a UE $D$ relative to the access point $A$ can be presented by its distance $r$ to $A$, and its azimuth and elevation AoA of signal $[\theta,\varphi]$. 

\section{3D Human Body Blockage Model}\label{sec:3d_hbb_model}
\subsection{Self-Blockage}
\begin{figure}[!h]
	\centering
	\begin{subfigure}[b]{.49\textwidth}
		\centering
		\includegraphics[width=.8\textwidth]{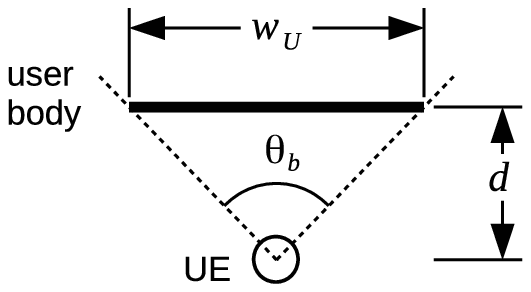}
		\caption{}
		\label{fig:horizontal_self_blocking_sector}
	\end{subfigure}
	\begin{subfigure}[b]{.49\textwidth}
		\centering
		\includegraphics[width=.9\textwidth]{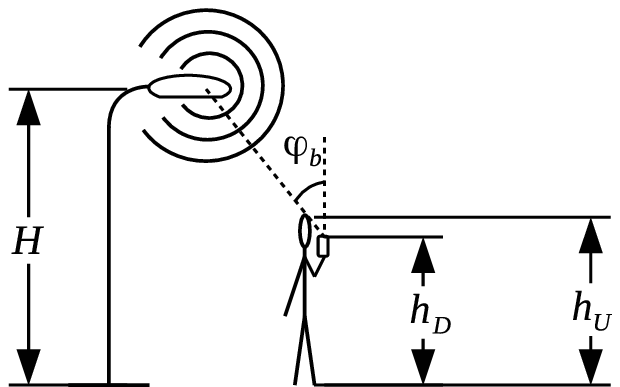}
		\caption{}
		\label{fig:vertical_self_blocking_sector}
	\end{subfigure}
	\caption{The (a) horizontal and (b) vertical self-blocking sectors can be defined by the azimuth and elevation AoAs of signal, respectively.}
\end{figure}

The very practical 2D self-blockage model proposed by \textit{Bai} and \textit{Heath, Jr.} \cite{bai2014analysis} can be briefly illustrated by Fig.\ref{fig:horizontal_self_blocking_sector}. The handset is hold in front of the user at a distance $d$, and the user's body is approximated as a flat blocker of width $w_\textrm{U}$. Both the transmitting and receiving antennas are considered to be infinitesimally-small with respect to the dimension of human body,  and therefore approximated as point antennas. A horizontal blocking sector is defined by
\begin{align}
|\theta|&<\frac{\theta_\textrm{b}}{2},\label{equ:horizontal_self_blockage_sector}\\
\theta_\textrm{b}&=2\arctan\frac{w_\textrm{U}}{2d},
\end{align}
where $\theta$ is the azimuth AoA of signal. Here, the impact of elevation angle of the transmitter to the receiver is ignored. This is a reasonable approximation for macro and small cells, where the horizontal distance between the UE and the base station is generally larger than the vertical distance with significance. However, for micro cell scenarios, the 2D approximation can deviate from the actual propagation in 3D space, especially when the UE is near the access point.

For a 3D extension, we assume that each user is $h_\textrm{U}$ tall and holding the handset by $h_\textrm{D}$ above the ground, while the access points are installed on the top of road poles with height of $H$, as shown in Fig. \ref{fig:vertical_self_blocking_sector}. In this way, a vertical blocking sector can be defined as
\begin{align}
\varphi&>\varphi_\textrm{b},\label{equ:vertical_self_blockage_sector}\\
\varphi_\textrm{b}&=\arctan\frac{d}{h_\textrm{U}-h_\textrm{D}},
\end{align}
where $\varphi$ is the elevation AoA of signal. A self-body blockage occurs if and only if the angle of arrival (AoA) of the signal falls in both the horizontal and vertical blocking sectors.

\subsection{Pedestrian Blockage}\label{subsec:pedestrian_blockage}
\begin{figure}[!h]
	\centering
	\begin{subfigure}[b]{.65\textwidth}
		\centering
		\includegraphics[width=\textwidth]{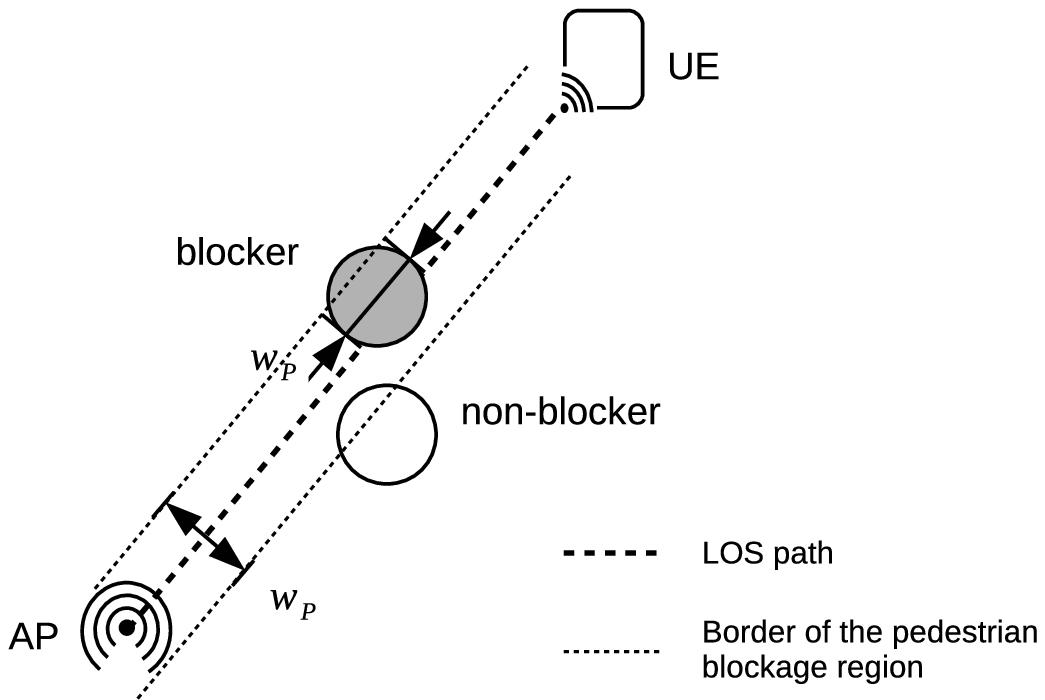}
		\caption{}
		\label{fig:2d_pedestrian_blockage}
	\end{subfigure}
	\begin{subfigure}[b]{.6\textwidth}
		\centering
		\includegraphics[width=\textwidth]{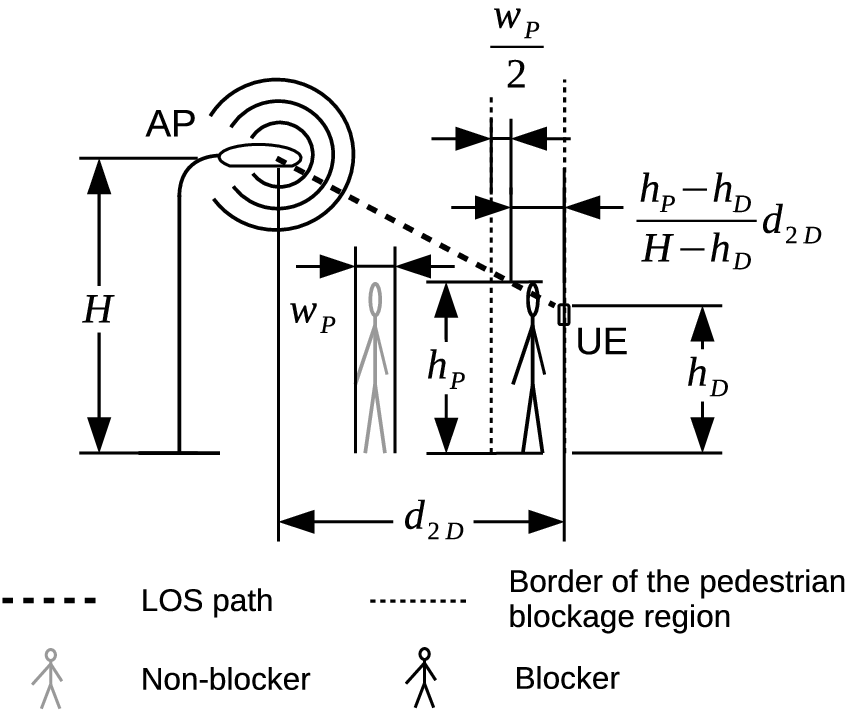}
		\caption{}
		\label{fig:3d_pedestrian_blockage}
	\end{subfigure}
	\caption{Given the UE position relative to the AP, we can define the pedestrian blockage region considering (a) the horizontal path and (b) the vertical path.}
\end{figure}

Pedestrians are usually modeled as 2D circles with certain diameter, which cause blockages when they intersect the path between the transmitter and the receiver \cite{venugopal2015analysis}, i.e. their centers fall into a $w_\textrm{P}$ wide blocking region, where $w_\textrm{P}$ is the pedestrian diameter, as illustrated in Fig. \ref{fig:2d_pedestrian_blockage}. Taking the elevation AoA into account, here we extend this model to a 3D version. Similar to the approach used in \cite{venugopal2016millimeter}, we model each pedestrian as a cylinder with diameter $w_\textrm{P}$ and height $h_\textrm{P}$. Once again, we ignore the dimension of antennas. As shown in Fig. \ref{fig:3d_pedestrian_blockage}, it can be calculated that a pedestrian cannot block the LOS path when its center is horizontally farther than $\frac{h_\textrm{P}-h_\textrm{D}}{H-h_\textrm{D}}\times d_\textrm{2D}+\frac{w_\textrm{P}}{2}$ away from the user, where $d_\textrm{2D}$ is the horizontal distance between the user and the AP.

\begin{figure}[!h]
	\centering
	\begin{subfigure}[b]{.7\textwidth}
		\includegraphics[width=\textwidth]{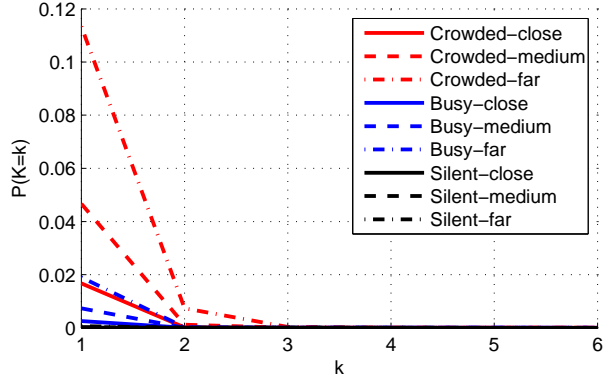}
		\caption{}
		\label{fig:pmf_pedestrian_num_3d}
	\end{subfigure}
	\begin{subfigure}[b]{.7\textwidth}
		\includegraphics[width=\textwidth]{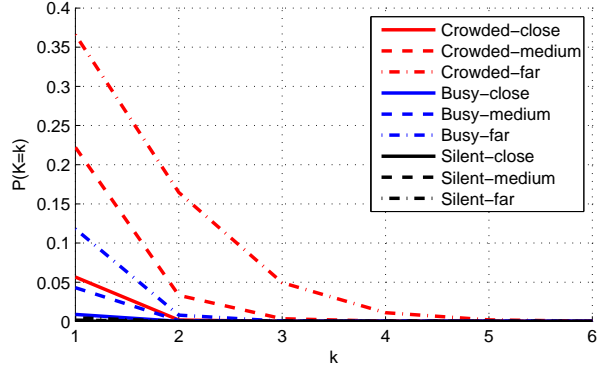}
		\caption{}
		\label{fig:pmf_pedestrian_num_2d}
	\end{subfigure}
	\caption{The probability mass function of arriving blocking pedestrians in a $T=\SI{0.1}{\second}$ time window under different reference scenarios and distances, calculated with (a) our 3D pedestrian blockage model and (b) the	2D pedestrian blockage model.}
	\label{fig:pmf_pedestrian_num}
\end{figure}

For simplification here we consider a single-user and micro-cell scenario, where only one AP is serving the cell and the only user is located horizontally $d_\textrm{2D}$ away from the AP. Pedestrians nearby are uniformly distributed in the area and moving at low speeds ($<3$ km/h) in independently binary uniformly distributed directions. Under these conditions we can assert that the amount of pedestrian blockages $K$ arriving in a certain time interval $\Delta t$ is Poisson distributed:
\begin{equation}\label{equ:poisson_distribution}
P(K=k,\Delta t) = \frac{(\lambda \Delta t)^ke^{-\lambda \Delta t}}{k!},
\end{equation}
where $\lambda$ is the expectation of $K$. Under the low mobility assumption, we can approximately consider $d_\textrm{2D}$ as constant when $t$ is small enough. As the positions of all pedestrians are independent and identically distributed (i.i.d.), we have
\begin{equation}\label{equ:pedestrian_density}
\lambda=\lambda_0\left(\frac{h_\textrm{P}-h_\textrm{D}}{H-h_\textrm{D}}d_\textrm{2D}+\frac{w_\textrm{P}}{2}\right)w_\textrm{P},
\end{equation}
where $\lambda_0$ is the average number of pedestrian arrivals in unit time and unit area in the cell.
To briefly investigate the pedestrian blockage rate in the outdoor scenario, we set $\Delta t=\SI{0.1}{\second}$, and the geometric parameters to realistic values as listed in Tab. \ref{tab:geo_params}. Note that we set $w_\textrm{U}=w_\textrm{P}$ and $h_\textrm{U}=h_\textrm{P}$, considering a common dimension of human bodies. But the parameters for user and pedestrians are distinguished from each other, in order to provide a flexibility for potential further analyses such as vehicle blockage or vehicle communication devices. We defined three different reference pedestrian scenarios, in which $\lambda_0=\SI{0.01}{\meter^{-2}\second^{-1}}$ (silent), $\SI{0.3}{\meter^{-2}\second^{-1}}$ (busy) and $\SI{2}{\meter^{-2}\second^{-1}}$ (crowded), as well as three different reference 2D distances $d_\textrm{2D}=\SI{15}{\meter}$ (far), \SI{5}{\meter} (medium) and \SI{1}{\meter} (close). The probability mass function (PMF) of $K$ under different reference pedestrian scenarios and user distances can be calculated according to (\ref{equ:poisson_distribution}) and (\ref{equ:pedestrian_density}), as shown in Fig. \ref{fig:pmf_pedestrian_num_3d}. It can be seen that the pedestrian blockage has a considerable impact under the busy and crowded scenarios, and the chance of $K\ge3$ within $\Delta t=\SI{0.1}{\second}$ is negligible. For comparison we also computed the PMF with the traditional 2D pedestrian blockage model, the result is presented in Fig.\ref{fig:pmf_pedestrian_num_2d}. Under all scenarios and at all distances, our 3D model gives a lower probability of pedestrian blockage than the 2D model, which reveals a higher potential of outdoor deployment of mmWave communication than expected before.

\begin{table}[!h]
	\centering
	\caption{Reference parameters in the studied outdoor scenario}
	\label{tab:geo_params}
	\begin{tabular}{c|c|c|c|c|c|c|c}
		Parameter & $w_\textrm{U}$ & $h_\textrm{U}$ & $d$ & $h_\textrm{D}$ & $h_\textrm{P}$ & $w_\textrm{P}$& $H$ \\\hline
		Value (\SI{}{\meter})  & 0.3 & 1.7 & 0.15 & 1.5 & 1.7 & 0.3 & 3
	\end{tabular}
\end{table}

The duration of a blockage depends on speeds and positions of both the user and the blocker. For simplification here we model it as uniformly distributed in the interval $[\tau_{\min},\tau_{\max}]$.

\section{Impact of Human Body Blockage on Transmission Efficiency}\label{sec:analysis}
\subsection{Sidewalk Scenario: a Case Study}
\begin{figure}[!h]
	\centering
	\includegraphics[width=.5\textwidth]{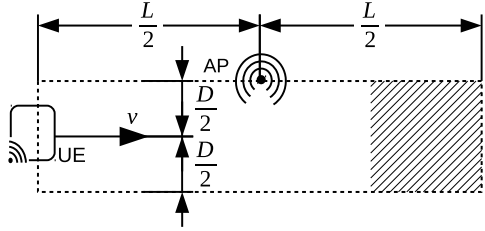}
	\caption{Top-view of the simplified single-user sidewalk scenario. The UE moves through the $L\times D$ cell area, the self-blocking region is illustrated with shadow.}
	\label{fig:simple_motion_model}
\end{figure}

To provide a rapid, intuitive and deep understanding in the impact of HBB on mmWave communication systems, simulations were carried out under a simplified use scenario. For the HBB problem, a multi-user case can be decomposed into several individual single-user multi-blocker cases, as long as the users have the same geometries as pedestrians. Hence, for simplification we investigated the case with a single user moving at a constant speed one-directionally along the middle axis of sidewalk, as illustrated in Fig. \ref{fig:simple_motion_model}. It can be seen that the self-blockage occurs only once, but lasts until the user leaves the cell, which is deterministic and easy to forecast through tracking. So first we focus on the pedestrian blockage. Given a frame $i$ in which the user is located at $[\theta_i,\varphi_i]$, the probability that at least one pedestrian blockage arrives in this frame can be calculated by
\begin{align}
P_i&=\sum\limits_{k=1}^\infty P(K_i=k,T)\nonumber\\
&=\sum\limits_{k=1}^\infty P(K=k,T|d_\textrm{2D}=(H-h_\textrm{D})\tan\varphi_i)\nonumber\\
&=\sum\limits_{k=1}^\infty\frac{(\lambda_iT)^ke^{-\lambda_iT}}{k!},
\label{equ:ped_blockage_prob}
\end{align}
where $\lambda_i=\lambda_0((h_\textrm{P}-h_\textrm{D})\tan\varphi_i+\frac{w_\textrm{P}}{2})w_\textrm{P}$.

Taking the uniformly distributed random duration of each blockage into account, the probability that no pedestrian blockage is present in a frame $i$ is that
\begin{equation}
\breve{P}_i=(1-P_i)\prod\limits_{j=i-\left\lceil\frac{\tau_{\max}}{T}\right\rceil}^{i-1}\sum\limits_{k=0}^\infty P(K_j=k)\left(\frac{(i-j)T-\tau_{\min}}{\tau_{\max}-\tau_{\min}}\right)^k.
\label{equ:ped_blockage_free_prob}
\end{equation}

Now take the self-blockage into account.
We denote the first frame at the cell entrance as $i=0$, the last frame before the self-blocking region entrance as $i=M-1$, and neglect the user movement during a frame for approximation. The UE position in an arbitrary frame $i$ can be presented by
\begin{align}
\theta_i &= \arctan\frac{D}{L-2viT},\\
\varphi_i &= \arctan\frac{\sqrt{D^2/4+(L/2-viT)^2}}{H-h_D}.
\end{align}

\subsection{Impacts of Pedestrian Scenario and Frame Length}
\begin{figure}[!h]
	\centering
	\begin{subfigure}[b]{.45\textwidth}
		\centering
		\includegraphics[width=\textwidth]{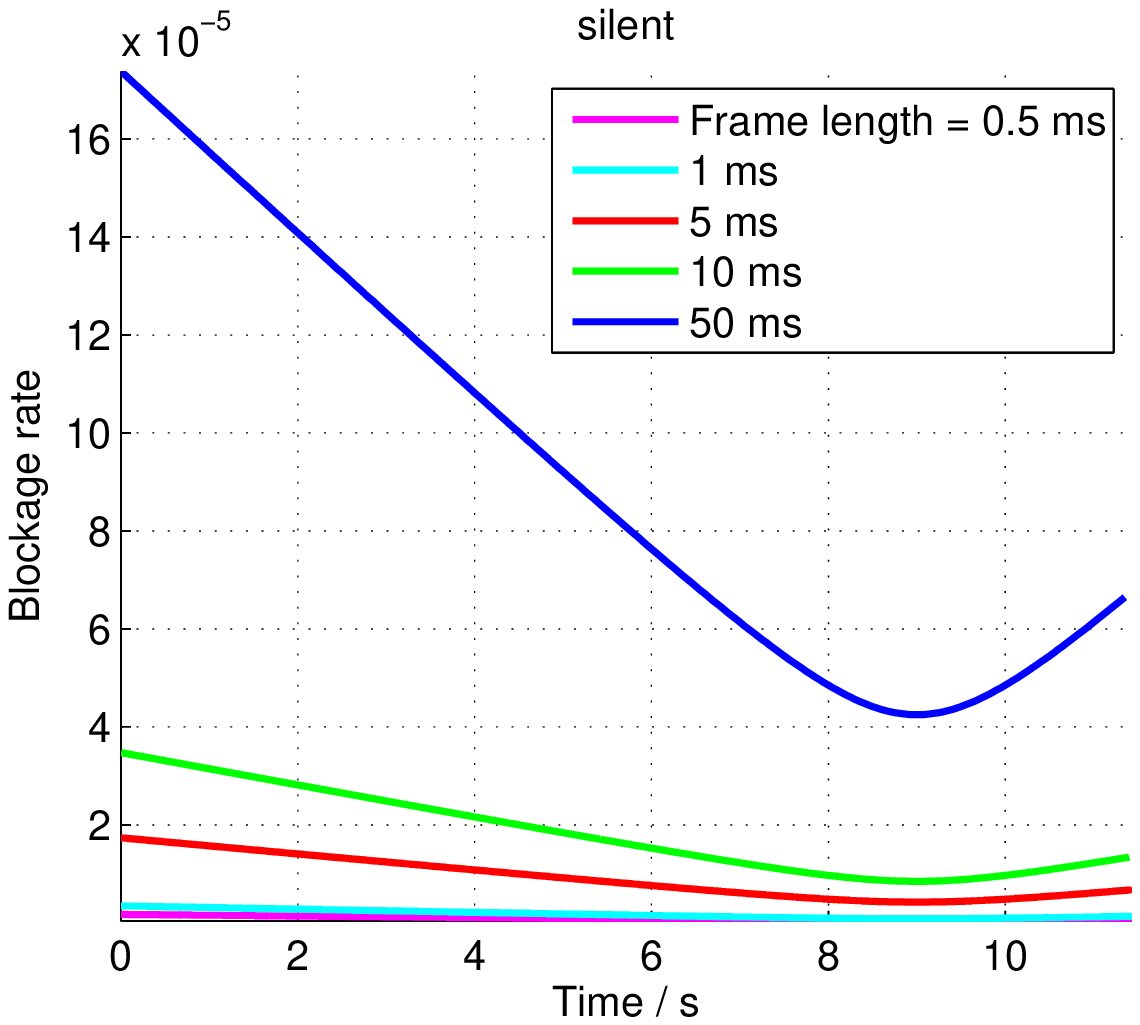}
		\caption{}
	\end{subfigure}
	\begin{subfigure}[b]{.45\textwidth}
		\centering
		\includegraphics[width=\textwidth]{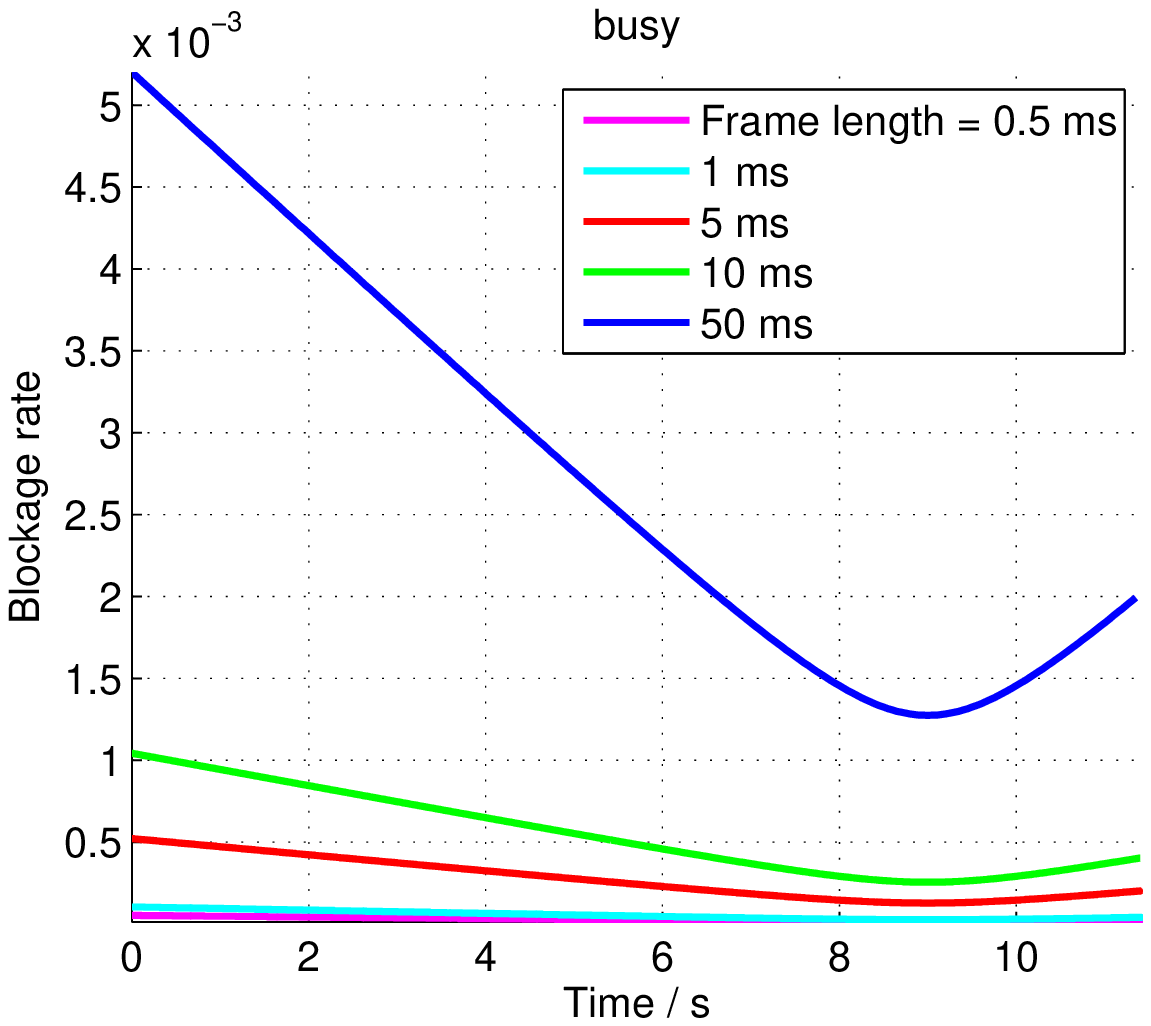}
		\caption{}
	\end{subfigure}
	\begin{subfigure}[b]{.45\textwidth}
		\centering
		\includegraphics[width=\textwidth]{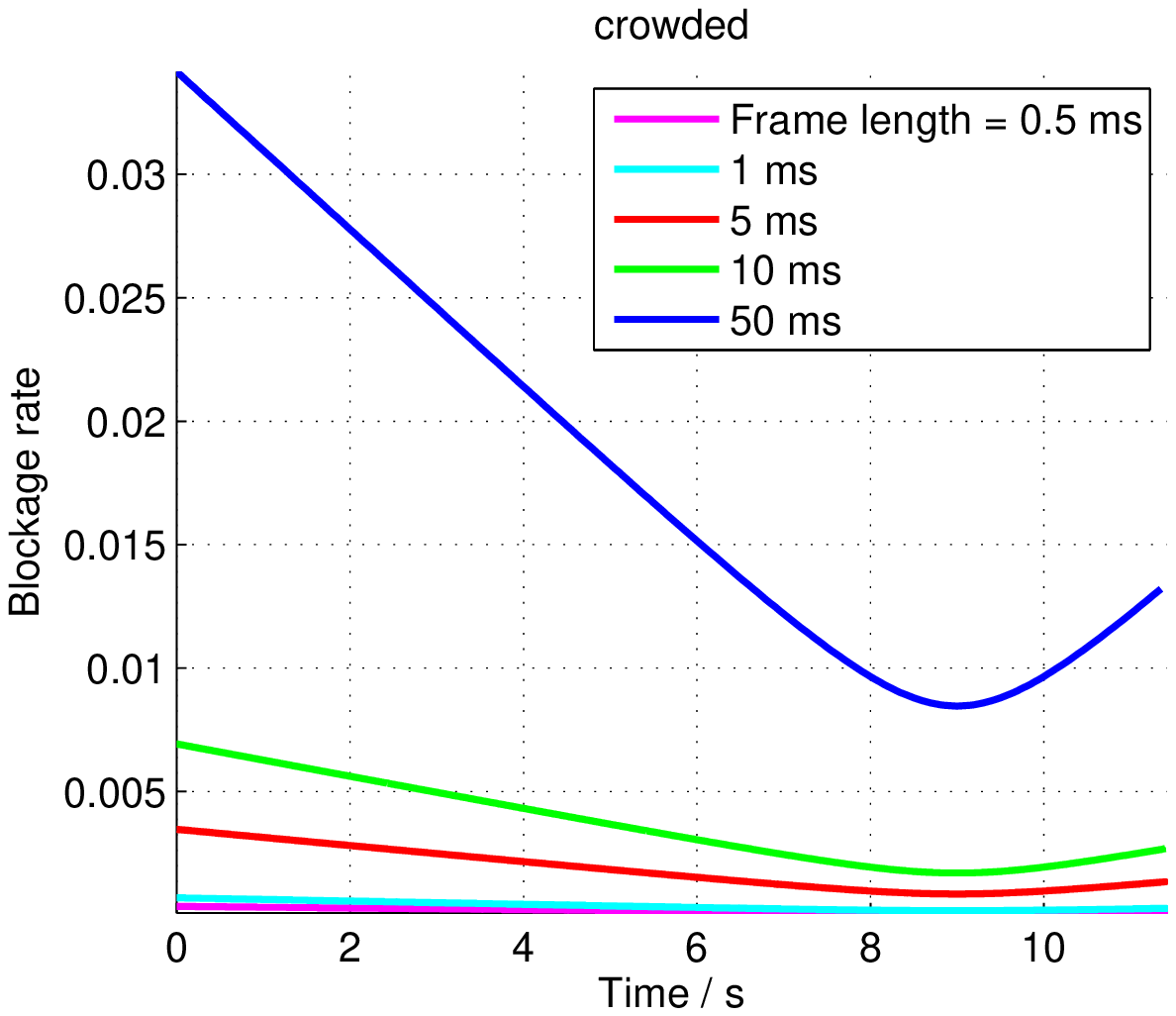}
		\caption{}
	\end{subfigure}
	\caption{The probability of blockage arrival in each frame under different frame lengths and different reference pedestrian scenarios: (a) silent, (b) busy and (c) crowded. Note that the total number of frames $M$ varies with the frame length $T$, although the total time remains almost the same. The asymmetry of the curves are caused by the self-blockage.}
	\label{fig:block_rate}
\end{figure}

For a brief evaluation we took the geometric parameter values in Tab.\ref{tab:geo_params}. We also set $L=\SI{15}{\meter}$, $D=\SI{2}{\meter}$ and $v=\SI{3}{\kilo\meter/\hour}$. Under this configuration, the user enters its self-blocking region at $[\theta=26.38^\circ,\varphi=36.87^\circ]$, and $M$ can be calculated for different values of frame length $T$:
\begin{equation}
	M = \left\lfloor\frac{L+D\cot 26.38^\circ}{2vT}\right\rfloor.
\end{equation}
Thus, according to (\ref{equ:ped_blockage_prob}), the blockage arrival probability of the $i^\textrm{th}$ frame $P_i$ is a function of $T$ and $\lambda_0$. We computed $P_i$ for the three reference pedestrian scenarios we defined in Sec. \ref{subsec:pedestrian_blockage}, the results are depicted in Fig. \ref{fig:block_rate}. A significant increase of the $P_i$ can be observed when $\lambda_0$ or $T$ grows. Additionally, this increase is more critical when the UE is relatively far from the AP, in comparison to the case that the UE is near the AP.

\begin{figure}[!h]
	\centering
	\begin{subfigure}[b]{.45\textwidth}
		\centering
		\includegraphics[width=\textwidth]{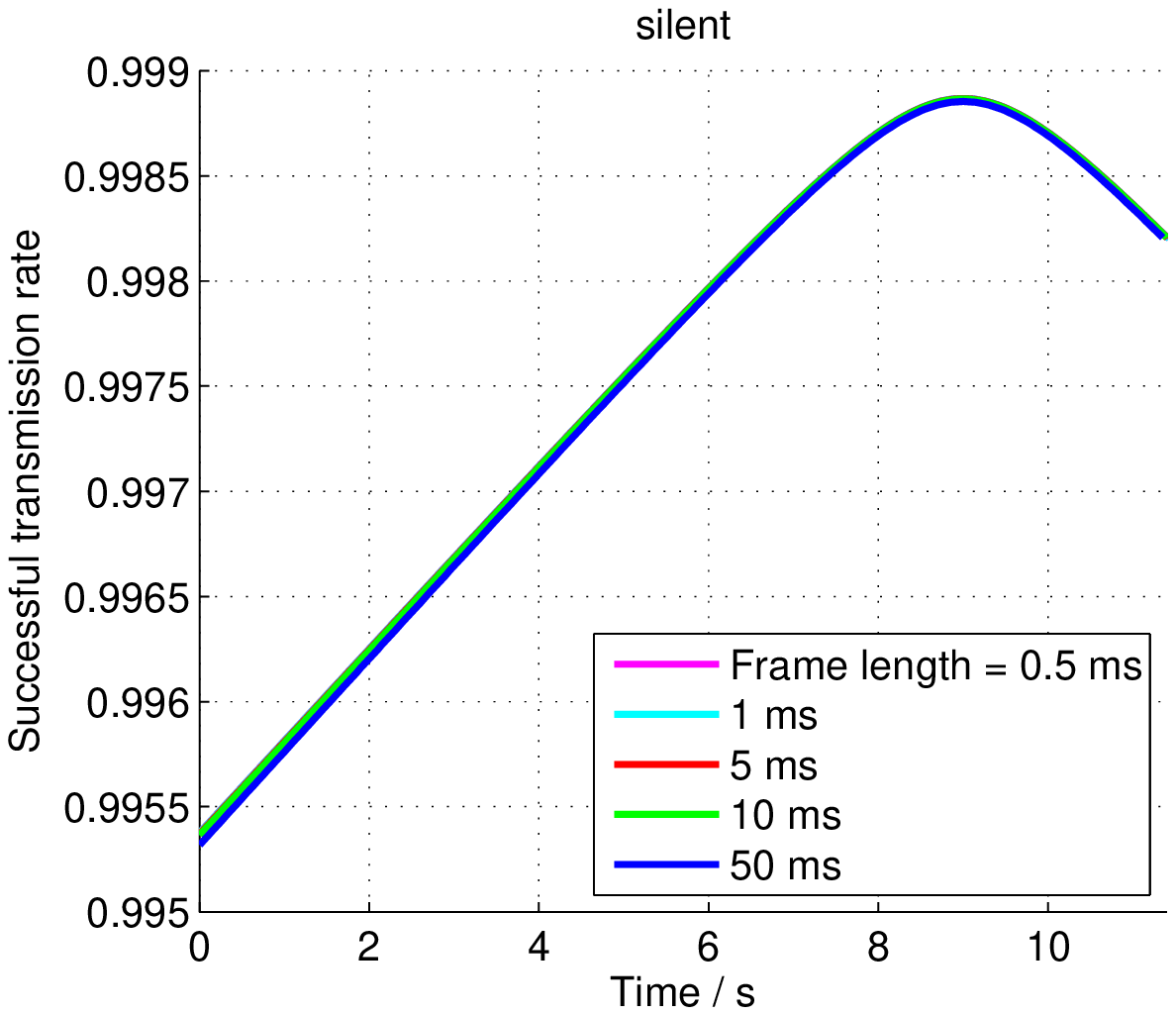}
		\caption{}
	\end{subfigure}
	\begin{subfigure}[b]{.45\textwidth}
		\centering
		\includegraphics[width=\textwidth]{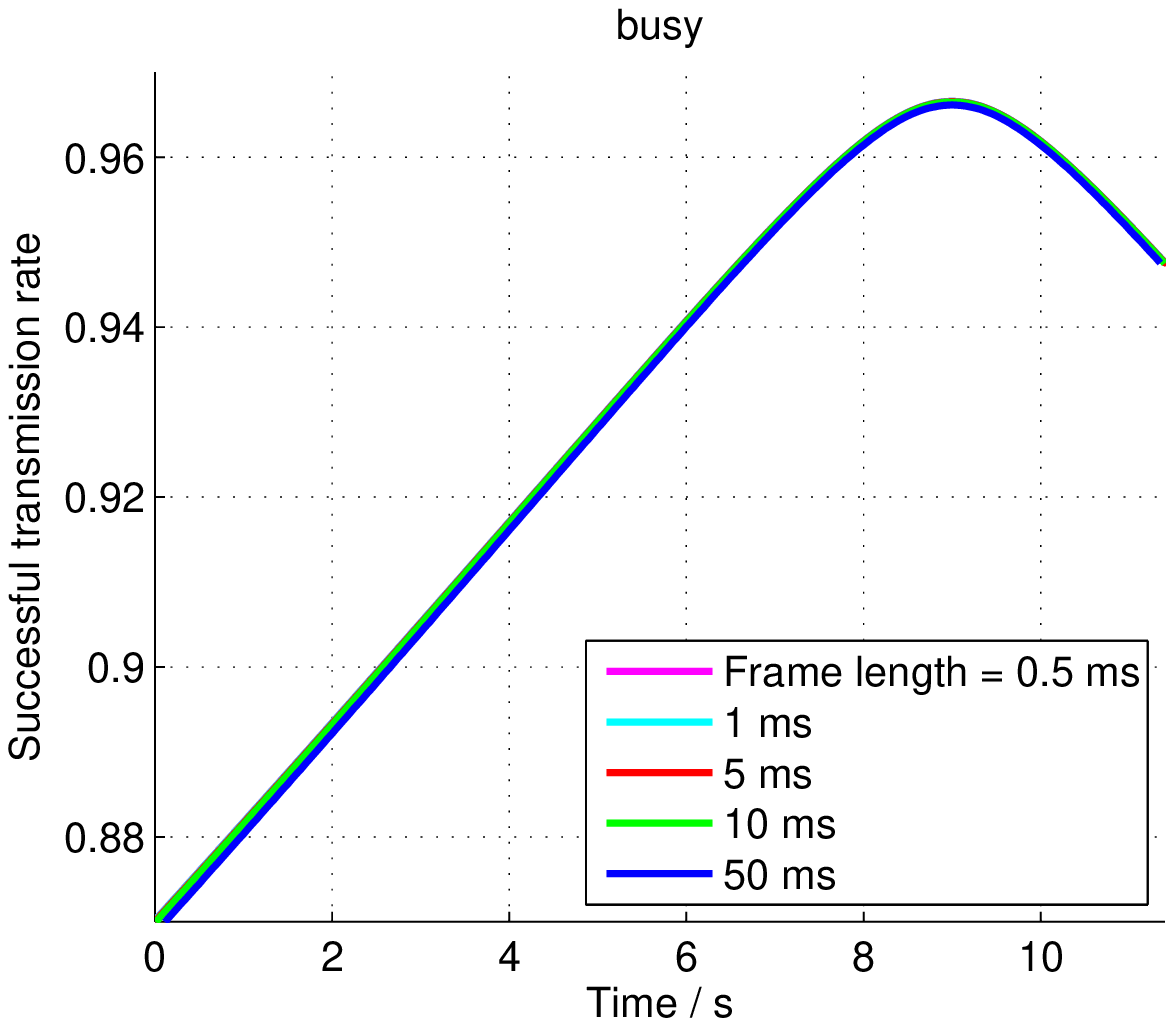}
		\caption{}
	\end{subfigure}
	\begin{subfigure}[b]{.45\textwidth}
		\centering
		\includegraphics[width=\textwidth]{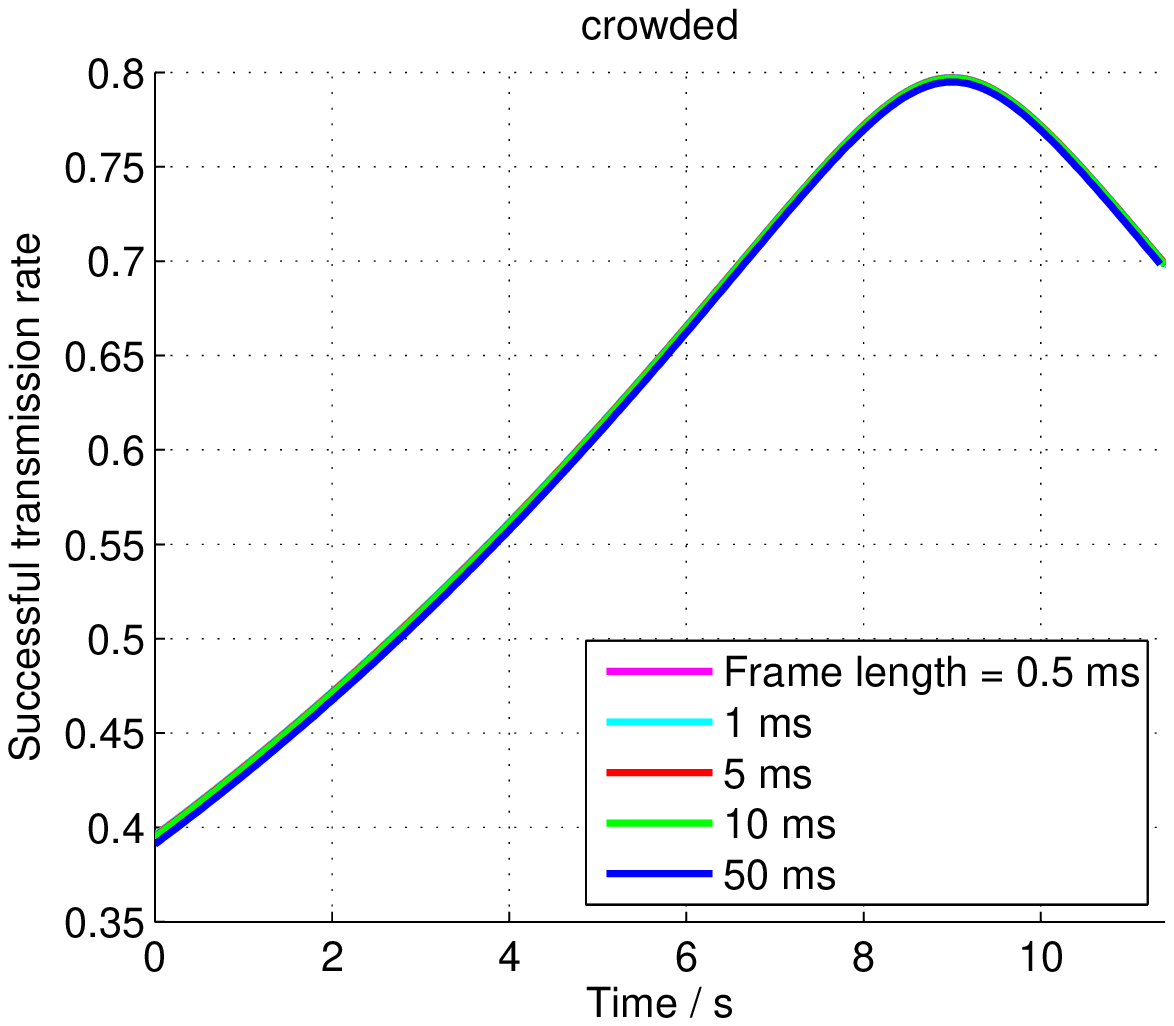}
		\caption{}
	\end{subfigure}
	\caption{The blockage-free probability in each frame under different frame lengths and different reference pedestrian scenarios: (a) silent, (b) busy and (c) crowded. Note that the total number of frames $M$ varies with the frame length $T$, although the total time remains almost the same. The asymmetry of the curves are caused by the self-blockage.}
	\label{fig:trans_rate}
\end{figure}

As the next step, according to \cite{collonge2004influence} we set $\tau_{\min}=\SI{0.5}{\second}$ and $\tau_{\max}=\SI{2}{\second}$. Thus, the probability of blockage-free in the $i^\textrm{th}$ frame $\breve{P}_i$ can also be computed with (\ref{equ:ped_blockage_free_prob}) as a function of $T$ and $\lambda_0$. Results under reference pedestrian scenarios are shown in Fig. \ref{fig:trans_rate}. Once again, a significant dependency on $\lambda_0$ can also be observed by $\breve{P}_i$. The frame blockage-free rate dramatically decreases when the pedestrian density grows. However, differing from the case of $P_i$, the frame length $T$ shows only a negligible impact on $\breve{P}_i$. This can be explained by the memorability of blockage events. As (\ref{equ:ped_blockage_free_prob}) shows, $\breve{P}_i$ relies on the historical blockage arrival status in the past period of $\tau_{\max}$ instead of only the current frame. As $\tau_{\max}\gg T$ and $\tau_{\max}\gg \Delta T$, there is 
\begin{equation}
	\left\lceil\frac{\tau_{\max}}{T}\right\rceil \simeq \left\lceil\frac{\tau_{\max}}{T+\Delta T}\right\rceil,
\end{equation}
and the impact of $\Delta T$ is hence canceled.

For a more intuitive and clearer evaluation we computed the HBB loss in \si{dB}. The pedestrian blockage loss was obtained from the expected frame transmission failure rate, and the self-blockage loss was obtained from the surface ratio of self-blocking region to the entire cell area. The results for different reference scenarios are listed in Tab. \ref{tab:hbb_loss_in_db}.

\begin{table}
	\centering
	\caption{HBB loss in \si{dB} under different reference scenarios}
	\begin{tabular}{c|c|c|c}
						&	Pedestrian blockage loss & Self-blockage loss & HBB loss \textsuperscript{*}\\\hline
		Silent		&	0.021	&	\multirow{3}{*}{3.953} & 3.974\\\cline{1-2}\cline{4-4}
		Busy		& 0.630		&	&	 4.583\\\cline{1-2}\cline{4-4}
		Crowded & 9.304		&	&13.257\\\hline
		\multicolumn{4}{l}{*: Here we consider the gross HBB loss of all transmission slots,}\\
		\multicolumn{4}{l}{including UL, DL and guarding interval slots.}
	\end{tabular}
	\label{tab:hbb_loss_in_db}
\end{table}

\begin{figure}[!h]
	\centering
	\includegraphics[width=.8\textwidth]{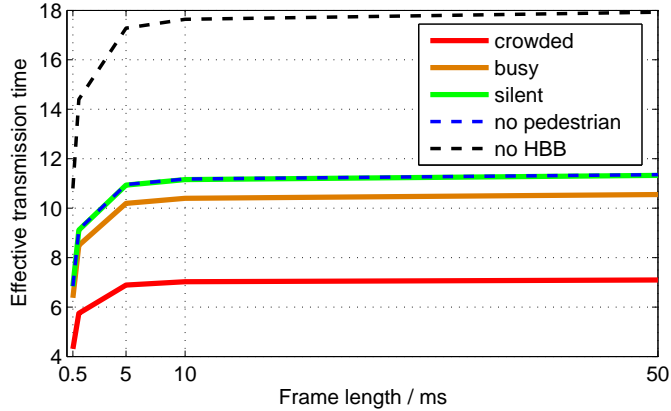}
	\caption{Expected effective DL transmission time as function of frame length}
	\label{fig:eff_trans_time}
\end{figure}

So far, a longer frame length leads to a higher blockage arrival probability, i.e. frame transmission failure rate. And once a frame fails, the data loss is also more with a longer frame length. Hence, the HBB-caused data loss increases with the frame length. Nevertheless, a longer frame can also benefit in some specific cases. For instance, if the system is designed for users to download large files from the cloud server, the data will mainly flow in DL, while only a short UL transmission slot is necessary to send reports. In this case, the expectation of overall effective DL transmission time during the stay of UE in the cell is of our interest, which can be calculated as
\begin{equation}
	\bar{t}_\textrm{data}=M\sum\limits_{i=0}^{M-1}\breve{P}_iT_1,
\end{equation}
which is proportional to $T_1$. If we fix $T_2$ and $T_3$ at a certain value e.g. one sTTI, so that the increase in $T$ completely contributes to $T_1$, which helps to increase $\bar{t}_\textrm{data}$. Meanwhile, $\breve{P}$ decreases with increasing $T$, which compensates this gain. To investigate the joint impact, we set $T_2=T_3=\SI{0.1}{\milli\second}$, and computed $\bar{t}_\textrm{data}$ for $T_1\in[\SI{0.3}{\milli\second},\SI{49.8}{\milli\second}]$, i.e. $T\in[\SI{0.5}{\milli\second},\SI{50}{\milli\second}]$. Three reference pedestrian scenarios were taken, along with two other cases, where
\begin{enumerate}
	\item no pedestrian exists;
	\item all types of HBB are ignored.
\end{enumerate}
The results are depicted in Fig. \ref{fig:eff_trans_time}.

Generally, $\bar{t}$ strictly increases with $T$, which means that the gain in $T_1$ overcomes the loss in $\breve{P}$ in all scenarios. This increase slows down as $T$ grows. Comparing the scenarios, it can be seen that HBB can be almost ignored in the silent scenario, but very critical in crowded scenario. Also we can assert that the self-blockage plays an important role, even more critical than the pedestrian blockages. Additionally, it worths to note that only HBB-caused loss is discussed here. In practice, a long frame length can also lead to an outdated CSI estimation, and therefore a system mismatch. This may generate extra transmission loss, which may overwhelm the gain in $\bar{t}$. Hence, the channel fast fading has to be taken into account to optimize $T$.

\subsection{Impact of Access Point Height}
Besides the pedestrian scenario and time frame specification, the deployment of access points, especially the AP height $H$, also plays a important role in the HBB phenomenon. First, the self-blocking region is jointly set by (\ref{equ:horizontal_self_blockage_sector}) and (\ref{equ:vertical_self_blockage_sector}). For any given $\theta$, changing $H$ leads to a change of $\phi$, and may hence influence the self-blockage state. Second, from (\ref{equ:poisson_distribution}) and (\ref{equ:pedestrian_density}) we know that increasing $H$ will decrease the probability of pedestrian blockage arrival in each frame.

\begin{figure}[!h]
	\centering
	\begin{subfigure}[b]{.45\textwidth}
		\centering
		\includegraphics[width=\textwidth]{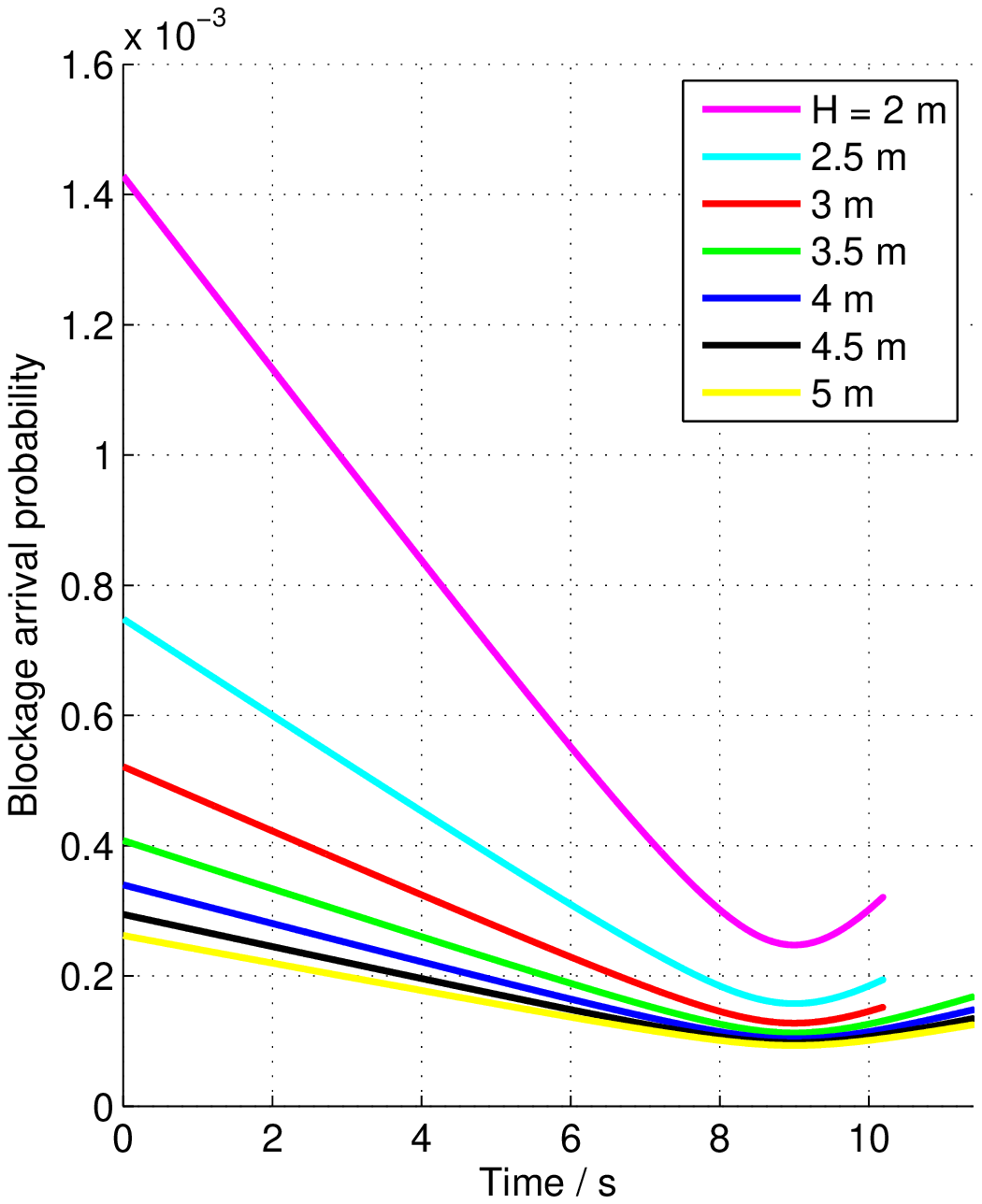}
		\caption{Blockage arrival probability}
		\label{fig:block_rate_ap_height}
	\end{subfigure}
	\begin{subfigure}[b]{.45\textwidth}
		\centering
		\includegraphics[width=\textwidth]{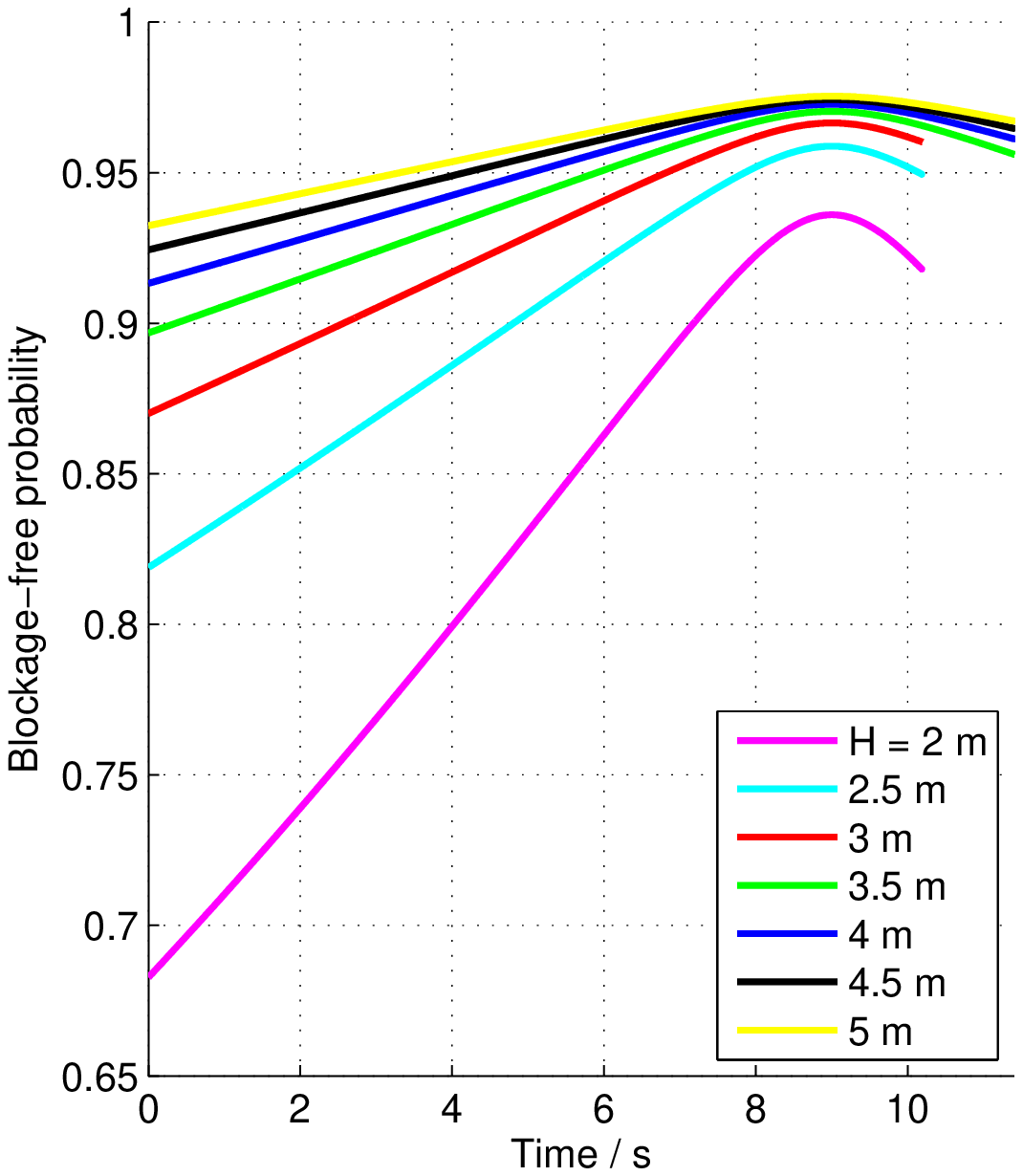}
		\caption{Blockage-freel probability}
		\label{fig:trans_rate_ap_height}
	\end{subfigure}
	\begin{subfigure}[b]{.65\textwidth}
		\centering
		\includegraphics[width=\textwidth]{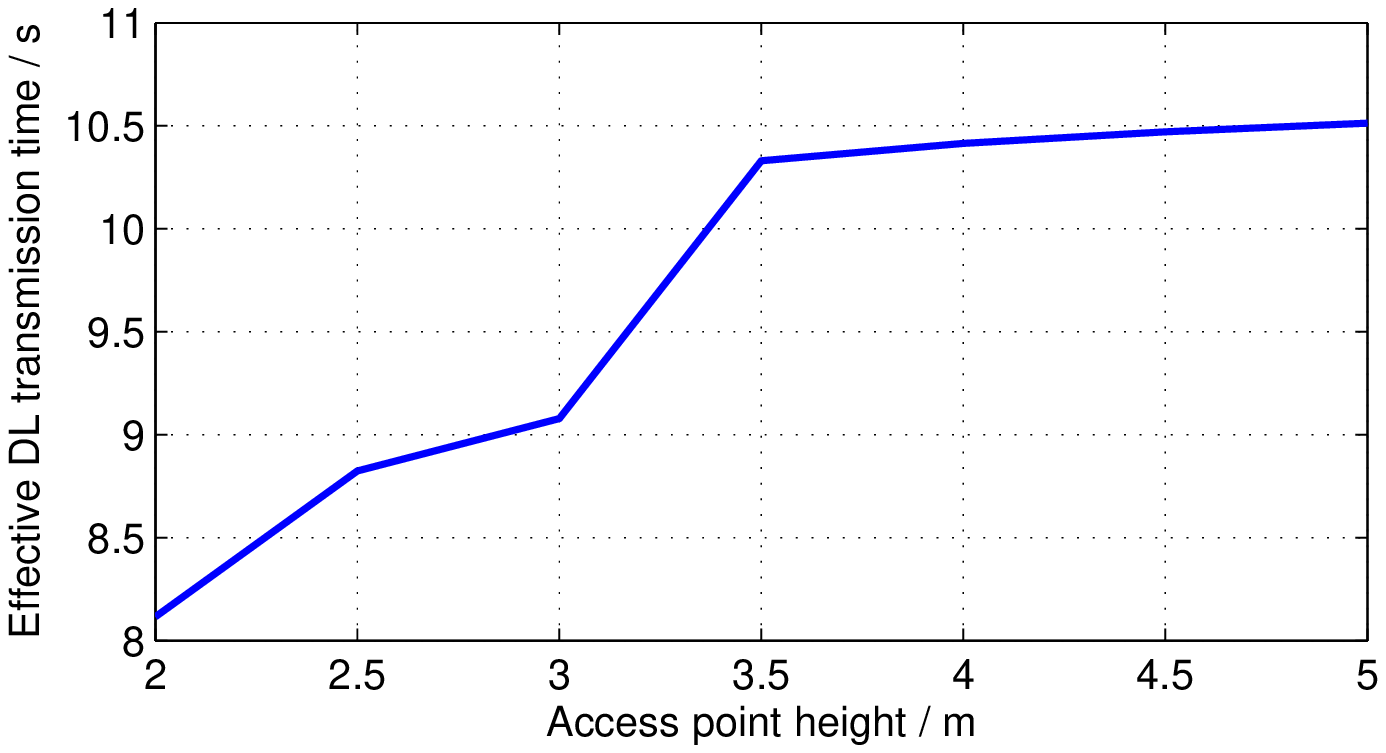}
		\caption{Effective DL transmission time}
		\label{fig:trans_time_ap_height}
	\end{subfigure}
	\caption{Impact of the AP height $H$ on HBB in frame-based transmission. In (a) and (b), the varying time range of different curves is caused by different self-blocking regions.}
	\label{fig:impact_ap_height}
\end{figure}

To demonstrate these effects, we calculated $P_i$, $\breve{P}_i$ and $\bar{t}$ with different values of $H$ from \SIrange{2}{5}{\meter}. The pedestrian scenario was set to \textit{busy}, frame length $T=\SI{5}{\milli\second}$, DL transmission slot length $T_1=\SI{4.8}{\milli\second}$, and the other geometric parameters as listed in Tab. \ref{tab:geo_params}. The results are shown in Fig. \ref{fig:impact_ap_height}. 

It can be significantly observed in Figs. \ref{fig:block_rate_ap_height} and \ref{fig:trans_rate_ap_height} that the self-blocking region varies with $H$. For $H\le\SI{3}{\meter}$, this region is determined by the vertical self-blocking sector; for $H\ge\SI{3.5}{\meter}$, it is determined by the horizontal self-blocking sector. Generally, a large $H$ is beneficial for avoiding HBB. Nevertheless, it also leads to a higher distance between AP and UE, which implies a stronger path loss, which must be considered in real system design.

\subsection{Impact of Human Body Size}
\begin{figure}[!h]
	\centering
	\begin{subfigure}[b]{.45\textwidth}
		\centering
		\includegraphics[width=\textwidth]{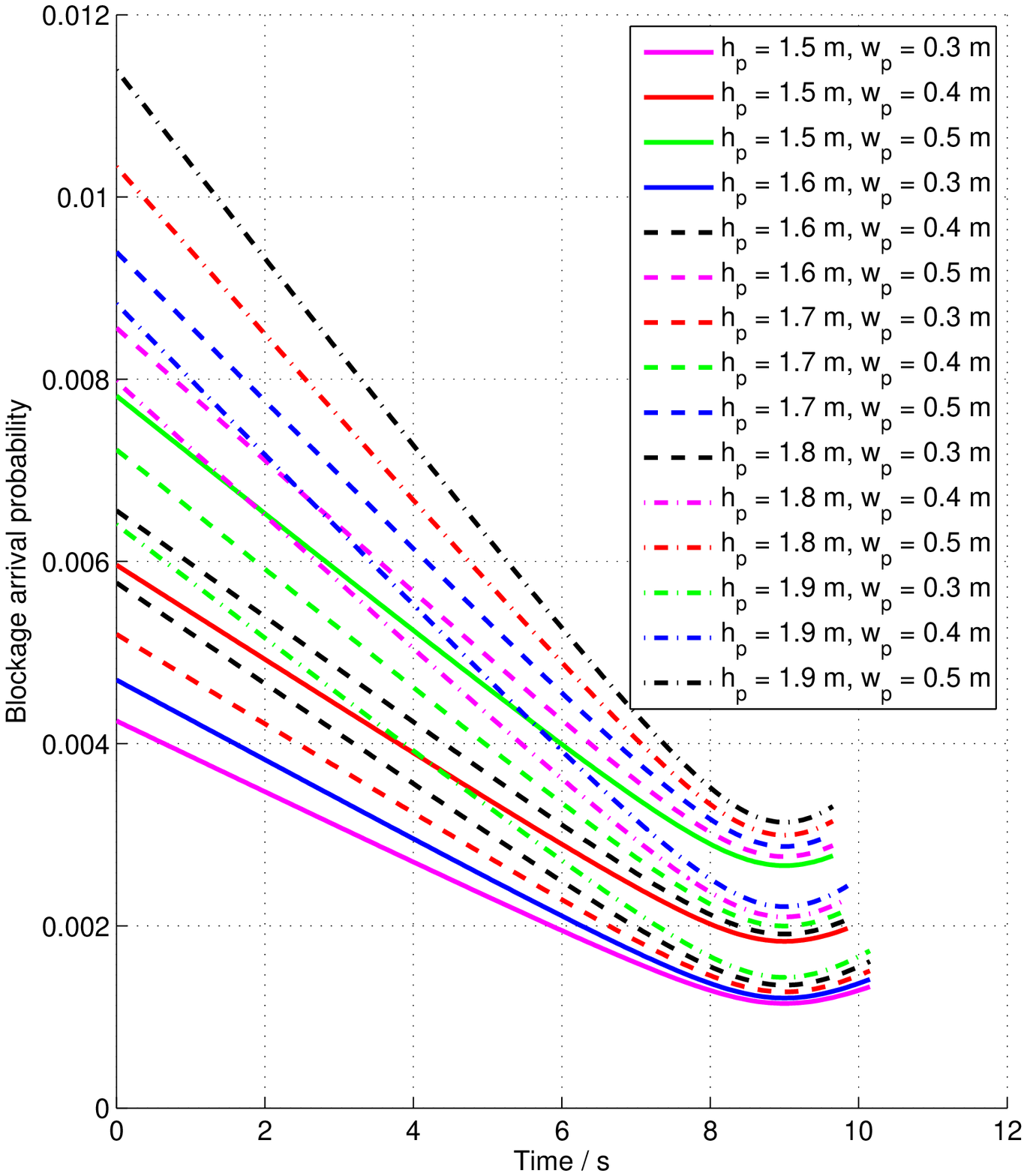}
		\caption{Blockage arrival probability}
		\label{fig:block_rate_human_body_geometry}
	\end{subfigure}
	\begin{subfigure}[b]{.45\textwidth}
		\centering
		\includegraphics[width=\textwidth]{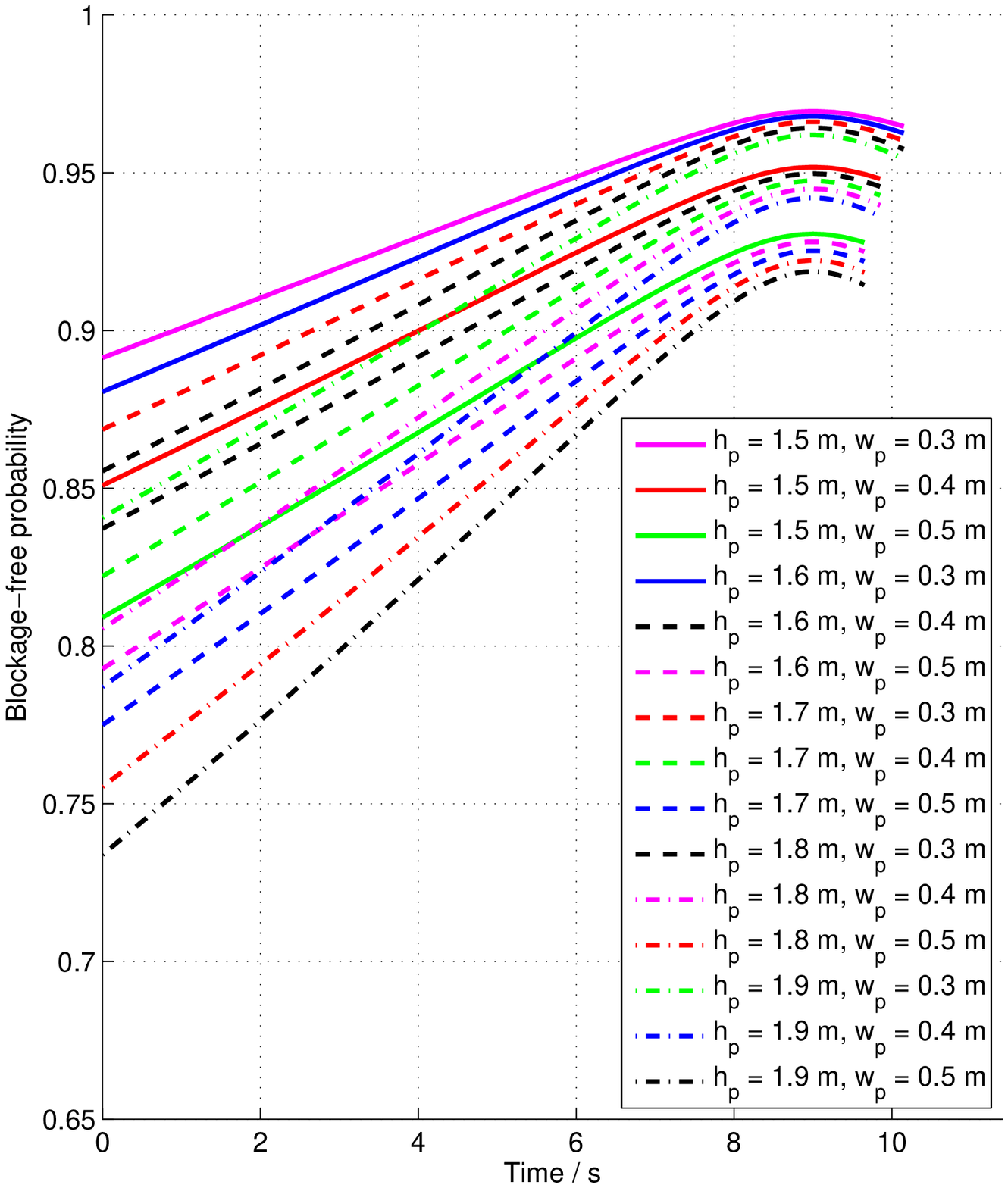}
		\caption{Blockage-freel probability}
		\label{fig:trans_rate_human_body_geometry}
	\end{subfigure}
	\begin{subfigure}[b]{.65\textwidth}
		\centering
		\includegraphics[width=\textwidth]{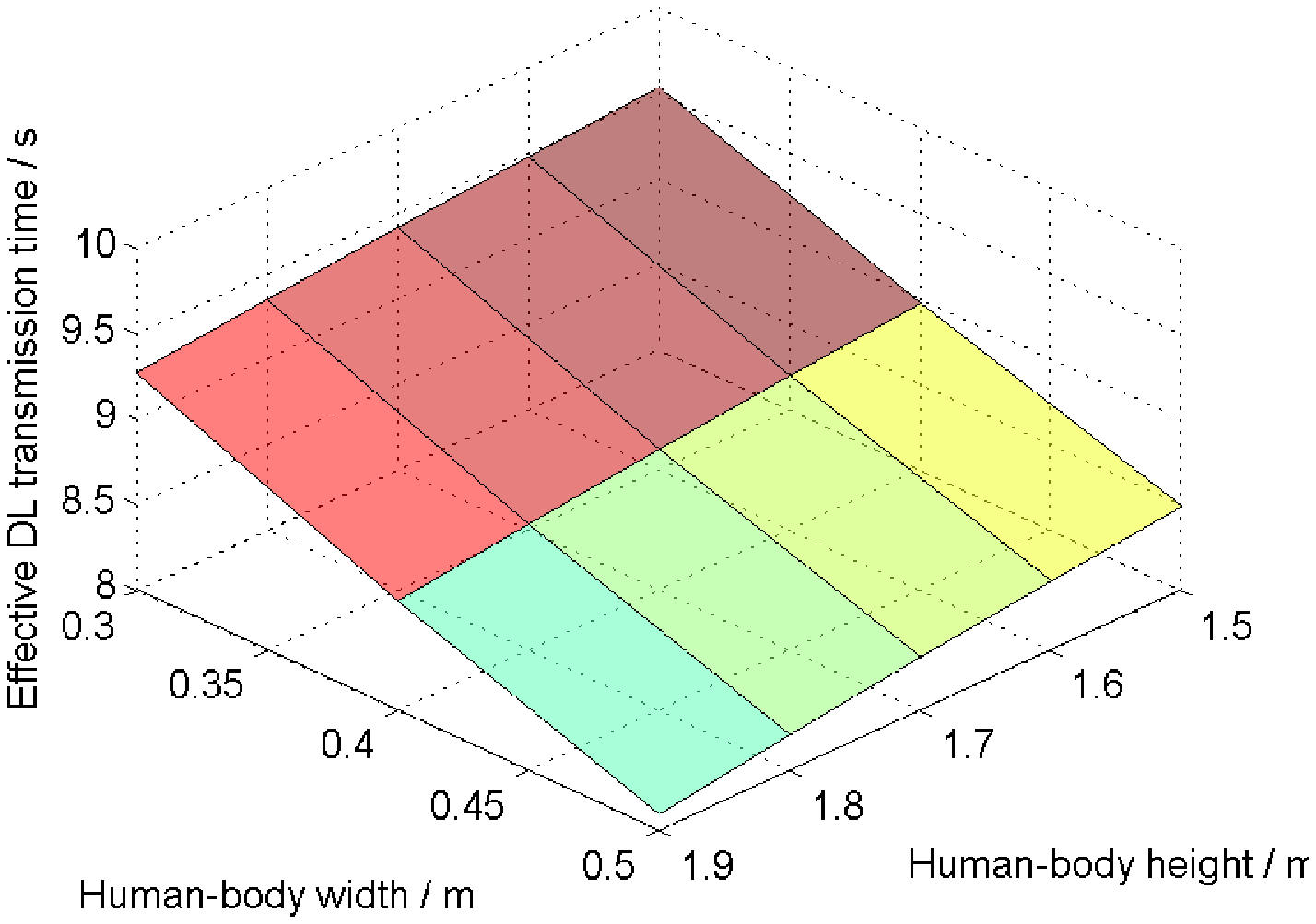}
		\caption{Effective DL transmission time}
		\label{fig:trans_time_human_body_geometry}
	\end{subfigure}
	\caption{Impact of the human-body size on HBB in frame-based transmission. In (a) and (b), the varying time range of different curves is caused by different self-blocking regions.}
	\label{fig:impact_human_body_geometry}
\end{figure}

Furthermore, the size of human bodies, including both the user and the pedestrians, also have impacts on the blocking range and blockage loss. To investigate these effects, we repeated the simulation of sidewalk scenario with varying human-body geometrical parameters. We take the values of $H=\SI{3}{\meter}$, $d=\SI{0.15}{\meter}$, $L=\SI{15}{\meter}$, $D=\SI{2}{\meter}$, $v=\SI{3}{\kilo\meter/\hour}$, $T_1=\SI{4.8}{\milli\second}$ and $T=\SI{5}{\milli\second}$. We consider a  homogeneous size of human bodies independent of the role (user or pedestrian) so that we tested $w_U = w_P$ in the range from \SIrange{0.3}{0.5}{\meter}, and $h_U=h_P$ in the range from \SIrange{1.5}{1.9}{\meter}. The device height is considered as proportional to the user height so that $h_D=\frac{1.5}{1.7}h_U$. As the simulation results in Fig. \ref{fig:impact_human_body_geometry} shows, the HBB loss generally raises with increasing height and width of human bodies. 


\section{Conclusion and Outlooks}\label{sec:conclusion}
In this paper, we have proposed a novel 3D physical model of human body blockage in outdoor mmWave communications. Compared to traditional models, our model takes the vertical drop between AP and UE into account,  and gives different results in the behaviors of HBB phenomena, including self-blockage and pedestrian blockage. Through case studies in a simplified single-user sidewalk downlink scenario, we have discussed the impacts of pedestrian density, frame length and access point height on the LOS transmission characteristics. Numerical computations have been conducted under different configurations, showing quantitative results.

For future work of ours and any other interested peers, there are still plenty of issues remaining open on this topic. First, more complex deployment scenario with multiple UEs and APs should be studied, in order to investigate the potential of practical system implementation. Second, the LOS channel availability should be analyzed in combination with path loss and channel fading, to obtain the overall mmWave channel quality. Furthermore, for future 5G mmWave cellular systems, our research has implied a potential of real-time system optimization with respect to the pedestrian density, which also worths further research.


\section*{References}

\begin{thebibliography}{19}
	\expandafter\ifx\csname natexlab\endcsname\relax\def\natexlab#1{#1}\fi
	\providecommand{\bibinfo}[2]{#2}
	\ifx\xfnm\relax \def\xfnm[#1]{\unskip,\space#1}\fi
	\bibitem[{Gupta and Jha(2015)}]{gupta2015survey}
	\bibinfo{author}{A.~Gupta}, \bibinfo{author}{R.~K. Jha},
	\newblock \bibinfo{title}{{A Survey of 5G Network: Architecture and Emerging
			Technologies}},
	\newblock \bibinfo{journal}{IEEE Access} \bibinfo{volume}{3}
	(\bibinfo{year}{2015}) \bibinfo{pages}{1206--1232}.
	\bibitem[{Soldani and Manzalini(2015)}]{soldani2015horizon}
	\bibinfo{author}{D.~Soldani}, \bibinfo{author}{A.~Manzalini},
	\newblock \bibinfo{title}{{Horizon 2020 and Beyond: on the 5G Operating System
			for a True Digital Society}},
	\newblock \bibinfo{journal}{IEEE Vehicular Technology Magazine}
	\bibinfo{volume}{10} (\bibinfo{year}{2015}) \bibinfo{pages}{32--42}.
	\bibitem[{Pi and Khan(2011)}]{pi2011introduction}
	\bibinfo{author}{Z.~Pi}, \bibinfo{author}{F.~Khan},
	\newblock \bibinfo{title}{{An Introduction to Millimeter-Wave Mobile Broadband
			Systems}},
	\newblock \bibinfo{journal}{IEEE Communications Magazine} \bibinfo{volume}{49}
	(\bibinfo{year}{2011}).
	\bibitem[{Niu et~al.(2015)Niu, Li, Jin, Su, and Vasilakos}]{niu2015survey}
	\bibinfo{author}{Y.~Niu}, \bibinfo{author}{Y.~Li}, \bibinfo{author}{D.~Jin},
	\bibinfo{author}{L.~Su}, \bibinfo{author}{A.~V. Vasilakos},
	\newblock \bibinfo{title}{{A Survey of Millimeter Wave Communications (mmWave)
			for 5G: Opportunities and Challenges}},
	\newblock \bibinfo{journal}{Wireless Networks} \bibinfo{volume}{21}
	(\bibinfo{year}{2015}) \bibinfo{pages}{2657--2676}.
	\bibitem[{Andrews et~al.(2017)Andrews, Bai, Kulkarni, Alkhateeb, Gupta, and
		Heath}]{andrews2017modeling}
	\bibinfo{author}{J.~G. Andrews}, \bibinfo{author}{T.~Bai},
	\bibinfo{author}{M.~N. Kulkarni}, \bibinfo{author}{A.~Alkhateeb},
	\bibinfo{author}{A.~K. Gupta}, \bibinfo{author}{R.~W. Heath},
	\newblock \bibinfo{title}{{Modeling and Analyzing Millimeter Wave Cellular
			Systems}},
	\newblock \bibinfo{journal}{IEEE Transactions on Communications}
	\bibinfo{volume}{65} (\bibinfo{year}{2017}) \bibinfo{pages}{403--430}.
	\bibitem[{3gp(2017)}]{3gpp2017further}
	\bibinfo{title}{{Further Advancements for E-UTRA Physical Layer Aspects
			(Release 9)}}, \bibinfo{type}{Technical Report} \bibinfo{number}{TR 36.814},
	Third Generation Partnership Project (3GPP), \bibinfo{year}{2017}.
	\bibitem[{Haenggi(2014)}]{haenggi2014mean}
	\bibinfo{author}{M.~Haenggi},
	\newblock \bibinfo{title}{{The Mean Interference-to-Signal Ratio and Its Key
			Role in Cellular and Amorphous Networks}},
	\newblock \bibinfo{journal}{IEEE Wireless Communications Letters}
	\bibinfo{volume}{3} (\bibinfo{year}{2014}) \bibinfo{pages}{597--600}.
	\bibitem[{Bai and Heath(2015)}]{bai2015coverage}
	\bibinfo{author}{T.~Bai}, \bibinfo{author}{R.~W. Heath},
	\newblock \bibinfo{title}{{Coverage and Rate Analysis for Millimeter-Wave
			Cellular Networks}},
	\newblock \bibinfo{journal}{IEEE Transactions on Wireless Communications}
	\bibinfo{volume}{14} (\bibinfo{year}{2015}) \bibinfo{pages}{1100--1114}.
	\bibitem[{Baccelli and Zhang(2015)}]{baccelli2015correlated}
	\bibinfo{author}{F.~Baccelli}, \bibinfo{author}{X.~Zhang},
	\newblock \bibinfo{title}{{A Correlated Shadowing Model for Urban Wireless
			Networks}},
	\newblock in: \bibinfo{booktitle}{2015 IEEE Conference on Computer
		Communications (INFOCOM)}, \bibinfo{organization}{IEEE}, pp.
	\bibinfo{pages}{801--809}.
	\bibitem[{Collonge et~al.(2004)Collonge, Zaharia, and
		Zein}]{collonge2004influence}
	\bibinfo{author}{S.~Collonge}, \bibinfo{author}{G.~Zaharia},
	\bibinfo{author}{G.~E. Zein},
	\newblock \bibinfo{title}{{Influence of the Human Activity on Wide-Band
			Characteristics of the 60 GHz Indoor Radio Channel}},
	\newblock \bibinfo{journal}{IEEE Transactions on Wireless Communications}
	\bibinfo{volume}{3} (\bibinfo{year}{2004}) \bibinfo{pages}{2396--2406}.
	\bibitem[{Lu et~al.(2012)Lu, Steinbach, Cabrol, and Pietraski}]{lu2012modeling}
	\bibinfo{author}{J.~S. Lu}, \bibinfo{author}{D.~Steinbach},
	\bibinfo{author}{P.~Cabrol}, \bibinfo{author}{P.~Pietraski},
	\newblock \bibinfo{title}{{Modeling Human Blockers in Millimeter Wave Radio
			Links}},
	\newblock \bibinfo{journal}{ZTE Communications} \bibinfo{volume}{10}
	(\bibinfo{year}{2012}) \bibinfo{pages}{23--28}.
	\bibitem[{Rajagopal et~al.(2012)Rajagopal, Abu-Surra, and
		Malmirchegini}]{rajagopal2012channel}
	\bibinfo{author}{S.~Rajagopal}, \bibinfo{author}{S.~Abu-Surra},
	\bibinfo{author}{M.~Malmirchegini},
	\newblock \bibinfo{title}{{Channel Feasibility for Outdoor Non-Line-of-Sight
			mmWave Mobile Communication}},
	\newblock in: \bibinfo{booktitle}{, 2012 IEEE Vehicular Technology Conference
		(VTC Fall)}, \bibinfo{organization}{IEEE}, pp. \bibinfo{pages}{1--6}.
	\bibitem[{Bai and Heath(2014)}]{bai2014analysis}
	\bibinfo{author}{T.~Bai}, \bibinfo{author}{R.~W. Heath},
	\newblock \bibinfo{title}{{Analysis of Self-Body Blocking Effects in Millimeter
			Wave Cellular Networks}},
	\newblock in: \bibinfo{booktitle}{48th Asilomar Conference on Signals, Systems
		and Computers}, \bibinfo{organization}{IEEE}, pp.
	\bibinfo{pages}{1921--1925}.
	\bibitem[{Venugopal et~al.(2015)Venugopal, Valenti, and
		Heath}]{venugopal2015analysis}
	\bibinfo{author}{K.~Venugopal}, \bibinfo{author}{M.~C. Valenti},
	\bibinfo{author}{R.~W. Heath},
	\newblock \bibinfo{title}{{Analysis of Millimeter Wave Networked Wearables in
			Crowded Environments}},
	\newblock in: \bibinfo{booktitle}{49th Asilomar Conference on Signals, Systems
		and Computers}, \bibinfo{organization}{IEEE}, pp. \bibinfo{pages}{872--876}.
	\bibitem[{Venugopal and Heath(2016)}]{venugopal2016millimeter}
	\bibinfo{author}{K.~Venugopal}, \bibinfo{author}{R.~W. Heath},
	\newblock \bibinfo{title}{{Millimeter Wave Networked Wearables in Dense Indoor
			Environments}},
	\newblock \bibinfo{journal}{IEEE Access} \bibinfo{volume}{4}
	(\bibinfo{year}{2016}) \bibinfo{pages}{1205--1221}.
	\bibitem[{Gapeyenko et~al.(2016)Gapeyenko, Samuylov, Gerasimenko, Moltchanov,
		Singh, Aryafar, Yeh, Himayat, Andreev, and
		Koucheryavy}]{gapeyenko2016analysis}
	\bibinfo{author}{M.~Gapeyenko}, \bibinfo{author}{A.~Samuylov},
	\bibinfo{author}{M.~Gerasimenko}, \bibinfo{author}{D.~Moltchanov},
	\bibinfo{author}{S.~Singh}, \bibinfo{author}{E.~Aryafar},
	\bibinfo{author}{S.-p. Yeh}, \bibinfo{author}{N.~Himayat},
	\bibinfo{author}{S.~Andreev}, \bibinfo{author}{Y.~Koucheryavy},
	\newblock \bibinfo{title}{{Analysis of Human-Body Blockage in Urban
			Millimeter-Wave Cellular Communications}},
	\newblock in: \bibinfo{booktitle}{2016 IEEE International Conference on
		Communications (ICC)}, \bibinfo{organization}{IEEE}, pp.
	\bibinfo{pages}{1--7}.
	\bibitem[{3gp(2017)}]{3gpp2017study}
	\bibinfo{title}{{Study on Channel Model for Frequency Spectrum above 6 GHz
			(Release 14)}}, \bibinfo{type}{Technical Report} \bibinfo{number}{TR 38.900},
	Third Generation Partnership Project (3GPP), \bibinfo{year}{2017}.
	\bibitem[{{Huawei's contribution to 3GPP TSG RAN WG1 Meeting
			84}(2016)}]{huawei2016overview}
	\bibinfo{author}{{Huawei's contribution to 3GPP TSG RAN WG1 Meeting 84}},
	\bibinfo{title}{{Overview of Short TTI}}, \bibinfo{type}{Technical Report}
	\bibinfo{number}{R1-160291}, \bibinfo{year}{2016}.
	\bibitem[{{Ericsson's contribution to 3GPP TSG RAN WG1 Meeting
			84}(2016)}]{erricson2016system}
	\bibinfo{author}{{Ericsson's contribution to 3GPP TSG RAN WG1 Meeting 84}},
	\bibinfo{title}{{System Level Evaluation Results for TTI Shortening
			Techniques}}, \bibinfo{type}{Technical Report} \bibinfo{number}{R1-161167},
	\bibinfo{year}{2016}.
	
\end{thebibliography}

\end{document}